\documentclass[12pt,preprint]{aastex}
\newcommand{\siml}{\lower4pt \hbox{$\buildrel < \over \sim$}}
\newcommand{\simg}{\lower4pt \hbox{$\buildrel > \over \sim$}}

\def\Mesz{M\'esz\'aros~}
\begin{document}
\title{A Characteristic Dense Environment or Wind Signature in Prompt GRB Afterglows}
\author{Shiho Kobayashi$^{1,2}$, Peter M\'esz\'aros$^{1,2,3}$ and  
Bing Zhang$^1$}
\affil{ $^{1}$Dpt. of Astronomy \& Astrophysics, Pennsylvania State
University, University Park, PA 16802\\
$^{2}$Center for Gravitational Wave Physics, Pennsylvania State University,
University Park, PA 16802 \\
$^{3}$Institute for Advanced Study, Princeton, NJ 08540}
\begin{abstract}
We discuss the effects of synchrotron self-absorption in the prompt 
emission from the reverse shock of GRB afterglows in a dense environment, 
such as the wind of a stellar progenitor or a dense ISM in early galaxies.
We point out that, when synchrotron losses dominate over inverse 
Compton losses, the higher self-absorption frequency in a dense 
environment implies a bump in the reverse shock emission spectrum, 
which can result in a more complex optical/IR light curve
than previously thought.  This bump is prominent especially if
the burst ejecta is highly magnetized. In the opposite case of low 
magnetization, inverse Compton losses lead to a prompt X-ray flare. 
These effects give a possible new diagnostic for the magnetic energy 
density in the fireball, and for the presence of a dense environment. 
\end{abstract}

\section{Introduction}
\label{sec:intro}

Snapshot fits of the broadband spectrum of gamma-ray burst (GRB)
afterglows to standard forward shock models have been found, in many
cases, to be consistent with an external environment density  $n\lesssim
1$ cm$^{-3}$ typical of a dilute interstellar medium (ISM) which, to
first order, can be taken to be approximately independent of distance
from the burst (Frail et al 2001). In other cases, the forward shock is
better fitted with an external density  which depends on distance as
$\rho\propto R^{-2}$, typical of a stellar wind environment (Chevalier
\& Li 1999, 2000; Li \& Chevalier 2003). The two types of fits have been
critically analyzed by e.g., Panaitescu and Kumar (2002), the conclusion
being that at least some bursts may occur in high mass-loss winds, as
expected from massive progenitors. The parameters for such wind fits 
are uncertain, due to poorly known stellar mass loss rates.

In this Letter we show that observations of prompt optical/IR 
and/or X-ray emission attributable to reverse shock emission could 
constrain the GRB environment. The reverse shock emission tends to 
be in the regime where the electron cooling time is shorter than 
the dynamical expansion time. In high density environments, 
such as a stellar progenitor wind  or a dense ISM in early galaxies, 
the self-absorption (SA) frequency is much higher than in the normal 
ISM, and it could be higher than the typical injection peak frequency 
(Wu et al. 2003). 
Here we argue, from general radiative transfer considerations, 
that in such situations when the emission is in the fast cooling regime 
and the SA frequency is higher than the injection frequency, the SA 
frequency and its scaling are different from, and the flux at the SA 
frequency is appreciably larger than, what had been previously estimated. 
This implies a different light curve time behavior for the afterglow 
prompt flash in a dense environment. This is of significant interest, 
since observations of the SA frequency and the net flux could
provide constraints on the otherwise poorly known wind mass-loss 
rates of progenitors stars, or on the presence of a dense ISM.
These new features are expected to be pronounced if the inverse Compton
process does not play a dominant role. This could happen in the case of
highly magnetized fireball ejecta, whose presence has been suggested by
some recent studies (Zhang, Kobayashi \& \Mesz 2003; Kumar \& Panaitescu
2003) and by reports of high gamma-ray polarization (Coburn \& Boggs
2003). The strength of the SA features would then provide a constraint
on the magnetization parameter.

\section{The Model}

We consider a relativistic shell (fireball ejecta) with an isotropic
energy $E$, an initial Lorentz factor $\eta$ and an initial width
$\Delta_0$ expanding into a surrounding medium with a density distribution
$\rho=A R^{-2}$.  The shell width $\Delta_0$ is related to the
intrinsic duration of the GRB as $\Delta_0 \sim (1+z)^{-1} cT$
(Kobayashi, Piran \& Sari 1997) where $z$ is the redshift of the burst. 
The interaction between the shell and the wind is described by two
shocks: a forward shock propagating into the wind and a reverse shock
propagating into the shell. The shocks accelerate electrons in the
shell and in the wind material, and the electrons emit photons via
synchrotron process. 

The evolution of the reverse shocks can be classified into two cases
depending on the value of the initial bulk Lorentz factor $\eta$ relative to 
a critical value $\eta_{cr}=((1+z)E/4\pi Ac^3T)^{1/4}$
(Sari \& Piran 1995; Kobayashi \& Zhang 2003b). If $\eta > \eta_{cr}$
(thick-shell case), the reverse shock starts out relativistic and 
it significantly decelerates the shell material, to $\Gamma \sim \eta_{cr}$. 
If $\eta < \eta_{cr}$ (thin-shell case), the reverse shock is initially 
Newtonian and becomes only mildly relativistic traversing the shell.
Most bursts in a wind environment fall in the thick-shell class,
($\eta > \eta_{cr}$), 
because the high density implies a critical Lorentz factor 
$\eta_{cr} \sim 70 ~\zeta^{1/4}E_{53}^{1/4}A_{11.7}^{-1/4} 
T_{50}^{-1/4}$ lower than the typical value in fireball models 
$\eta \gtrsim 100$ (e.g., Lithwick \& Sari 2001),
where  $\zeta=(1+z)/2$, $E_{53}=E/10^{53}$ ergs, the duration
$T_{50}=T/50$ sec, and $A_{11.7}=A/5\times 10^{11}$ g cm$^{-1}$ is a
typical wind mass loss rate. 
We are especially interested in long bursts with $t_\times \gtrsim 1$ 
minute, for the purpose of analyzing the light curve behavior of the reverse 
shock emission. We focus on the thick-shell case, for which one can take 
the shock crossing time $t_\times\sim T$ and the bulk Lorentz factor during 
the shock crossing ($t<t_\times$) is $\Gamma\sim\eta_{cr}$. 

If we neglect synchrotron self-absorption (which is discussed in 
the next section) the spectrum of the reverse shock is described by a 
broken power law with a peak $F_{\nu, max}$  and two break frequencies:
a typical frequency $\nu_{m}$ and a cooling frequency $\nu_{c}$ (Sari,
Piran \& Narayan 1998). Assuming that constant fractions ($\epsilon_e$
and $\epsilon_B$) of the internal energy produced by the shock go into
the electrons and  the magnetic field, the reverse shock spectrum at 
a shock crossing time $t_\times \sim T$ is characterized by 
(Kobayashi \& Zhang 2003b),
\begin{eqnarray}
\nu_{c}(T) &\sim&  
1.5\times10^{11} ~\zeta^{-3/2} \epsilon_{B,-1}^{-3/2} 
E_{53}^{1/2}A_{11.7}^{-2} T_{50}^{1/2}~\mbox{Hz}
\label{eq:nucr} \\
\nu_m(T) &\sim& 
5.0\times10^{12} ~\zeta^{-1/2} \epsilon_{e,-2}^2 \epsilon_{B,-1}^{1/2}
E_{53}^{-1/2}A_{11.7} \eta_2^2 T_{50}^{-1/2} ~\mbox{Hz}
\label{eq:numr} \\
F_{\nu,max}(T) &\sim& 
95 ~d^{-2} \zeta^2 \epsilon_{B,-1}^{1/2}
E_{53}A_{11.7}^{1/2} \eta_2^{-1} T_{50}^{-1}  ~\mbox{Jy}
\label{eq:Fr}
\end{eqnarray}
where $\epsilon_{B,-1}=\epsilon_B/10^{-1}$,  
$\epsilon_{e,-2}=\epsilon_e/10^{-2}$, 
$\eta_2=\eta/100$, $d=d_L(z)/(2\times10^{28}$cm), $d_L(z)$ 
is the luminosity distance of the burst, and 
$d_L(1)\sim 2\times 10^{28}$cm for the standard cosmological
parameters ($\Omega_m=0.3$, $\Omega_\Lambda=0.7$ and $h=0.7$).
The value of $\epsilon_B (\epsilon_e)$ assumed for the reverse 
shock emission are larger (smaller) than in some previous broadband 
fits of afterglow forward shocks (e.g. Wijers \& Galama 1999; Panaitescu 
\& Kumar 2001). However, the values of the equipartition parameters can 
in principle differ in the forward and reverse shock regions, and as 
shown recently, $\epsilon_B$ in the reverse (fireball ejecta) shock 
could be much higher than in the forward shock (Zhang, Kobayashi \& 
\Mesz 2003; Kumar \& Panaitescu 2003).

\section{Self-absorption and O/IR reverse flash}

In a stellar wind the external density at the initial interaction with 
the shell is much larger than in the typical ISM, hence the
cooling frequency $\nu_{c}$ is lower than the typical peak frequency 
$\nu_m$, and the synchrotron self-absorption (SA) frequency $\nu_{a}$ 
can be much higher than $\nu_m$: $\nu_{c}(T) < \nu_m(T) < \nu_{a}(T) $. 
As suggested by some recent work (Zhang, Kobayashi \& \Mesz 2003; Kumar 
\& Panaitescu 2003; Coburn \& Boggs 2003), the fireball ejecta could be 
highly magnetized. In such cases the condition $\epsilon_e/\epsilon_B \ll 1$ 
can lead to a Compton parameter $Y\ll 1$ and the inverse Compton process is 
not important for electron cooling (the opposite case $Y\gg 1$ is discussed 
below). A consequence of the synchrotron dominance is that self-absorption 
suppresses the emission below the SA frequency $\nu_a$, and prevents the 
electrons from cooling down to the Lorentz factor $\gamma_c$ corresponding 
to the cooling frequency $\nu_{c}$. The suppressed radiation energy is 
partly redistributed among the electrons, and results in a distinctive 
hump in the spectrum of the reverse shock emission.  

The reverse shock injects electrons with a power law energy distribution 
$N(\gamma)d\gamma \propto \gamma^{-p}d\gamma (\gamma \ge \gamma_m)$.
The energy deposited in electrons with Lorentz factors between $\gamma_m$
and $\gamma_a$ is $E_e \sim (p-1)N_t\Gamma \gamma_m m_e c^2/(p-2)$ where
$N_t$ is the number of electrons in the shell, $\gamma_a$ is the Lorentz
factor corresponding to the SA frequency, 
and we assumed $p>2$ and $\gamma_a \gg \gamma_m$ (so $E_e$ is 
essentially the total electron energy). This energy is 
redistributed among the electrons and photons in the optically thick 
regime on a timescale comparable to the cooling time (Ghisellini \&
Svensson 1989). Since the reverse shock is in the fast cooling regime
$(\gamma_c < \gamma_m < \gamma_a)$, the electrons have enough time to
redistribute the energy. In a dynamical time, the energy $E_e$ is
radiated as photons around $\nu_a$. Assuming $p=3$, the
flux at  $\nu_a$ is given by 
\begin{equation}
F_{\nu_a}(T) \sim \frac{1}{4\pi d_L^2} \frac{E_e}{\nu_a T}
\sim \frac{2(\nu_c\nu_m)^{1/2}}{\nu_a}F_{\nu,max}.
\label{eq:Fnua}
\end{equation}
A simple estimate of the maximal self-absorption flux is given by 
a black body flux with the reverse shock temperature (e.g. Sari \& 
Piran 1999; Chevalier \& Li 2000),
$F_{\nu_a}^{bb}=2\pi (1+z)^3  \nu_a^2 m_e  \Gamma  \gamma_{a} 
(R_\perp/d_L)^2$ where $R_\perp \sim 2 \Gamma c T/(1+z)$ is 
the observed size of the shell. Equating $F_{\nu_{a}}^{bb} \sim 
F_{\nu_a}$, we obtain the SA frequency 
\begin{equation}
\nu_{a}(T)  \sim 8.1 \times 10^{13} \zeta^{-3/14}\epsilon_{e,-2}^{2/7}
\epsilon_{B,-1}^{1/14}A_{11.7}^{2/7}E_{53}^{1/14} T_{50}^{-11/14} ~\mbox{Hz}.
\label{eq:nua}
\end{equation}
An alternative derivation of the SA frequency is obtained by requiring
the electron synchrotron cooling rate and heating rate (through absorption) 
to be equal at $\gamma_a$.  The cross section for the synchrotron absorption 
process is approximately $\sigma_{s} \sim \gamma^{-5} r_e r_L$
(e.g. Ghisellini \& Svensson 1991) where $r_e=q_e^2/m_ec^2$ and
$r_L=m_ec^2/q_eB$ are the classical electron radius and the Larmor
radius, $q_e$ is electron charge, $B=(32\pi \epsilon_B \Gamma^2 
\rho c^2)^{1/2}$ is comoving magnetic field strength behind the shock. 
Using this cross section and the photon flux determined by
eq. (\ref{eq:Fnua}), we can evaluate the heating rate, while the cooling
rate is given by the electron synchrotron power. Equating these rates 
reproduces the SA frequency of eq. (\ref{eq:nua}). 

\begin{figure}
\plotone{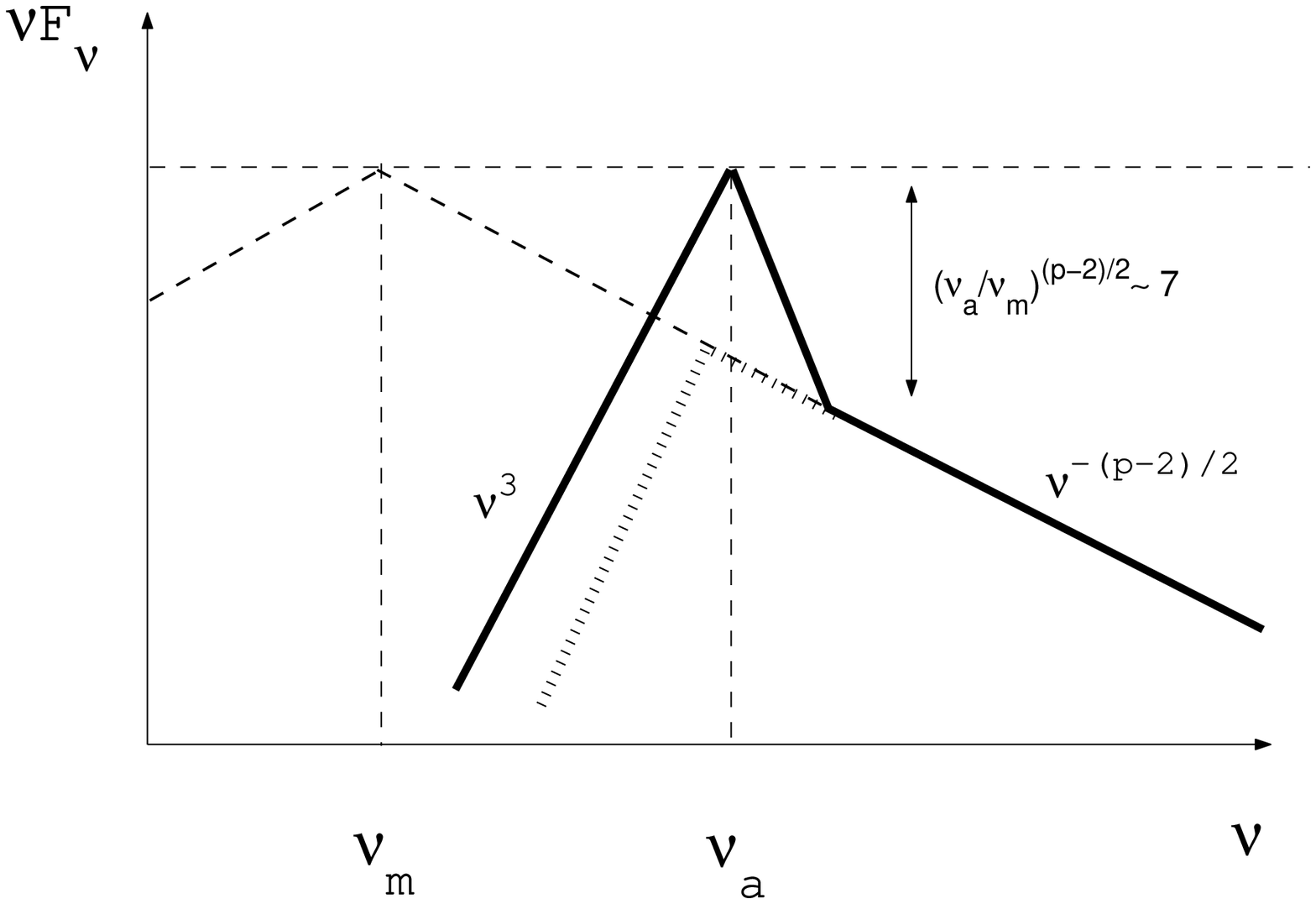}
\caption{Reverse shock spectrum in a dense environment or wind
  when synchrotron losses dominate: with self-absorption (thick solid) 
  and  without self-absorption (thin dashed).
  The schematic self-absorption maximum would appear as
  a rounded thermal peak. The previous self-absorbed flux
  estimate is shown by hashed lines.
  The correction factor $2(\nu_{a}/\nu_c)^{(p-2)/2}\propto t^{(p-2)/7}$
  is slightly larger at later times ($t \sim t_\times$), the value of
  $\sim 7$ is evaluated at $t_{a}$ for typical parameters.
 \label{fig1}}
 \end{figure}

Following Kobayashi \& Zhang (2003b), we obtain the 
scalings at $t<t_\times$ of the spectral quantities
$\nu_{c} \propto t, \nu_{m} \propto t^{-1},  \nu_{a}\propto t^{-5/7},
F_{\nu,max} \propto t^0$.  Although $\nu_{c}$ itself is not observed, as
a result of the  self-absorption, the flux at $\nu \gtrsim \nu_{a}$ can
be still estimated from these scalings, because the electron
distribution $N_e$ producing some observed frequency $\nu$ (say in the
K-band $\nu_K >\nu_{a}$) (and frequencies above this) is determined by
the distribution of injected electrons at the shock, and by the
synchrotron radiation cooling. Therefore, even though the flux at
$\nu_{a}$ has a hump which was previously unaccounted for, we can apply
the conventional synchrotron model to estimate the light curve at a
frequency $\nu \gtrsim \nu_{a}$. The optical/IR luminosity initially
increases as $\sim (\nu/\nu_{a})^2F_{\nu_a} \propto  t^{15/7}$. When the
SA frequency passes through the observation band $\nu_{obs}$ at
$t_{a}$, the flux reaches a peak of 
$F_{\nu_{obs}}(t_{a}) \sim 2(\nu_m\nu_c)^{1/2}
F_{\nu,max}/\nu_{obs}$ and then it rapidly decreases.
For electrons in the hump which are quasi-thermally distributed, 
the flux beyond the peak would decrease $\propto \exp(-t^{5/7})$
(or, if the emission of the quasi-thermal electrons above the peak is 
fitted by a power law $\nu^{-k}$ above the SA frequency, the 
decrease is $\propto t^{-(5/7)(k-1)}$), the flux dropping by a factor
$\mathcal{R}\sim 2 [\nu_{obs}/\nu_m(t_{a})]^{(p-2)/2}$. 
The time $t_a$ of the absorption peak passage through the K-band, 
the peak flux $F_{\nu_K}(t_{a})$ and the peak flux contrast $\mathcal{R}$ 
relative to the subsequent power-law decay value are given by
\begin{eqnarray}
&t_{a} &\sim  20 ~ \zeta^{-3/10}\epsilon_{e,-2}^{2/5}\epsilon_{B,-1}^{1/10}
A_{11.7}^{2/5}E_{53}^{1/10} T_{50}^{-1/10} \nu_{obs,14.2}^{-7/5} ~\mbox{sec},
\label{eq:tsa}\\
&F_{\nu_K}(t_{a}) &\sim 1 ~d^{-2}\zeta\epsilon_{e,-2} E_{53} T_{50}^{-1}
\nu_{obs,14.2}^{-1} ~\mbox{Jy},
\label{eq:fnukta}\\
&\mathcal{R} &\sim 7 ~\zeta^{1/10}\epsilon_{e,-2}^{-4/5}
\epsilon_{B,-1}^{-1/5}A_{11.7}^{-3/10}E_{53}^{3/10}
\eta_2^{-1}T_{50}^{-3/10} \nu_{obs,14.2}^{-1/5}
\end{eqnarray}
where $\nu_{obs,14.2} = \nu_{obs}/(1.6\times10^{14} \mbox{Hz})$ and 
$p=3$ was assumed. At this turnover, a color change from blue
to red is expected (see fig \ref{fig1}).  Since the polarization is zero 
for an optically thick quasi-thermal spectrum, the reverse shock shell 
can emit polarized photons only above the turnover. If a turnover 
characterized by $t_a, F_{\nu_K}(t_a)$ is observed, we can constrain
the mass loss rate $A$, assuming that the redshift $z$ is measured and 
the GRB explosion energy $E$ is determined, e.g. from late time bolometric 
afterglow observations. The peak flux determines the equipartition parameter 
$\epsilon_e$ via eq. (\ref{eq:fnukta}), and the peak time gives through 
eq. (\ref{eq:tsa}) a constraint on the mass loss rate,
\begin{equation} 
A \sim 5\times10^{11} \epsilon_{B,-1}^{-1/4}\zeta^{3/4}\epsilon_{e,-2}^{-1}
E_{53}^{-1/4} T_{50}^{1/4} \nu_{obs,14.2}^{7/2}(t_a/20\mbox{sec})^{5/2}~
\hbox{g~cm}^{-1}, 
\label{eq:massloss}
\end{equation}
where the dependence on the parameter $\epsilon_B$ is weak. 

\begin{figure}
\plotone{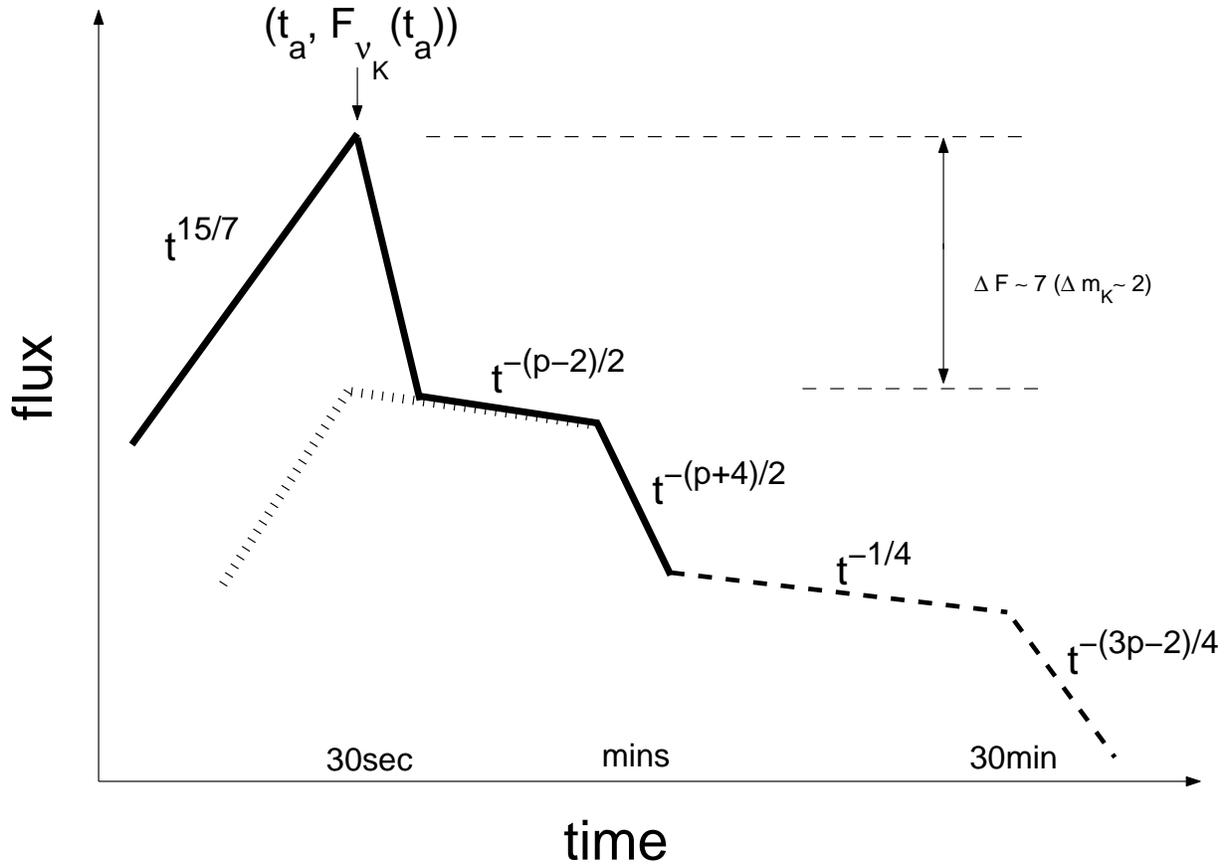}
\caption{Schematic optical light curve for a synchrotron
dominated fireball in a dense environment or wind:
reverse shock emission (solid) and forward shock emission (dashed).
The hashed line shows a previous estimate. Time scales are
rough estimates for the typical parameters.
 \label{fig2}}
 \end{figure}

After the optical/IR reverse shock emission drops to the level expected
from the usual synchrotron model, the flux decreases slowly as
$\sim (\nu_m/\nu_c)^{-1/2}(\nu/\nu_m)^{-p/2} F_{\nu,max} \sim t^{-(p-2)/2}$. 
Beyond a timescale comparable to the burst duration $T$, the 
optical/IR emission fades rapidly, because no further electrons are 
shocked in the shell (allowing the initially weaker but longer lasting 
forward shock to gradually dominate).
The angular time delay effect prevents the abrupt disappearance of the 
reverse component, whose flux decreases steeply as  $\sim t^{-(p+4)/2}$
(Kumar \& Panaitescu 2000; Kobayashi \&  Zhang 2003b). 
No color change is expected around this break. The flux at this break for 
$p=3$ is 
\begin{equation}
F_{\nu_K}(T) \sim 
0.1 ~d^{-2} \zeta^{3/4}\epsilon_{e,-2}^{2}\epsilon_{B,-1}^{1/4}
A_{11.7}^{1/2}E_{53}^{3/4} \eta_2 T_{50}^{-5/4} \nu_{obs,14.2}^{-3/2} 
~\mbox{Jy}.
\label{eq:FT}
\end{equation}
These two breaks are schematically shown in Fig. \ref{fig2}, the 
break sharpness being an idealization; in reality these would be rounded.
In our treatment, we have ignored pair formation ahead of the
blast wave, e.g. Beloborodov (2002), which may modify the 
forward shock spectrum and light curve.

\section{IC effects and prompt reverse X-ray flash}
\label{sec:xrflash}

The above discussion assumed magnetized ejecta with $\epsilon_e/\epsilon_B\ll 1$  
and $Y<1$. However, for weaker ejecta magnetization with $\epsilon_e/\epsilon_B 
\gg 1$, the inverse Compton (IC) process can affect the observed spectrum and 
the light curve, resulting in a reverse shock prompt X-ray flare. 
(For forward shocks, the importance of IC emission has been discussed, 
e.g. by Sari \& Esin 2001; Panaitescu \& Kumar 2001). 
Although the fraction of photons scattered to higher energies is small, 
they carry away a majority of the electron energy.
The energy available for the synchrotron process is reduced from the 
injected energy $E_e$ by a factor of $(1+Y)$ where
$Y\sim(\epsilon_e/\epsilon_B)^{1/2}$ is the Compton $Y$ parameter
(e.g. Zhang \& \Mesz 2001). The IC process enhances the electron
cooling, so $\nu_c$ is smaller by a factor of $(1+Y)^2$ than its
previous (synchrotron only) value. The thermal bump is shifted to a
lower frequency, becoming less prominent or even disappearing.
As a consequence of the IC cooling, the flux at $\nu_a$ becomes 
smaller by a factor of $\sim (1+Y)$ than the value given by 
eq. (\ref{eq:Fnua}). If $\nu_a$ is shifted below $\nu_m$, most of the
energy available for the synchrotron process is radiated between
$\nu_a$ and $\nu_m$. The flux at $\nu_a$ is reduced from
eq. (\ref{eq:Fnua}) by a factor of 
$\sim (1+Y) (\gamma_m/\gamma_a) \sim (1+Y) (\nu_m/\nu_a)^{1/2}$. By
equating the flux at $\nu_a$ and the black body flux at the reverse
shock characteristic temperature $\gamma_a \propto \nu_a^{1/2}$, we can 
obtain the SA frequency. For the same parameters as for
eq. (\ref{eq:Fnua})  but taking 
$\epsilon_{e,-1}=\epsilon_e/10^{-1}$ and 
$\epsilon_{B,-3}=\epsilon_B/10^{-3}$,
we get $\nu_a(T) \sim 5.8 \times 10^{13}$ 
$\zeta^{-3/14}\epsilon_{e,-1}^{1/7} \epsilon_{B,-3}^{3/14}$
$A_{11.7}^{2/7}E_{53}^{1/14} T_{50}^{-11/14}$  Hz and
$\nu_m(T) \sim 5.0\times10^{13} ~\zeta^{-1/2} \epsilon_{e,-1}^2
\epsilon_{B,-3}^{1/2} E_{53}^{-1/2}A_{11.7} \eta_2^2 T_{50}^{-1/2}$ Hz.
To produce a significant bump, the SA frequency $\nu_a$ should be much 
higher than the typical frequency $\nu_m$. When $\nu_a \lesssim \nu_m$,
as in this case, the contrast $\mathcal{R}$ is $\lesssim 2$ and the bump 
practically disappears.  The optical/IR light curve initially increases
as $\sim (\nu/\nu_a)^{5/2}F_{\nu_a} \propto t^{5/2}$, and after $\nu_a$ 
crosses the observation band, it decreases as $t^{-(p-2)/2}$. 
The transition is expected to be gradual and smooth. 

Since at the shock crossing time the forward and reverse shocked
regions have roughly comparable energy, the $\nu F_{\nu}$ peaks of the
synchrotron emissions reach roughly similar levels.  Assuming that the 
characteristic reverse shock IC frequency $\nu_{a,IC} \sim \gamma_a^2 
\nu_a$ is close to the typical (peak) frequency of the forward shock 
synchrotron emission, we can infer that in the case of $Y\ll 1$ the
reverse shock IC component is generally masked by the forward shock
synchrotron emission, whereas in the case of $Y\gg 1$ the reverse shock 
IC peak sticks out above the forward shock synchrotron peak.
Therefore, for weakly magnetized fireballs (with $Y \gtrsim 1$), 
a prompt X-ray flare is expected from the reverse shock. 
The characteristic photon energy and the flux at this
frequency are  $h\nu_{a, IC}(T) \sim 
20 ~\zeta^{-3/7}\epsilon_{e.-1}^{2/7} \epsilon_{B,-3}^{-1/14}
E_{53}^{1/7} A_{11.7}^{1/14}T_{50}^{-4/7}$ keV and 
$F_{\nu_a,IC}(T)\sim 1 ~d^{-2} \zeta^{31/14} \epsilon_{e,-1}^{9/7} 
\epsilon_{B,-3}^{3/7} E_{53}^{9/14} A_{11.7}^{15/14} 
T_{50}^{-17/14}~\hbox{mJy}$, respectively. The typical duration of this
X-ray flare at $\nu\gtrsim \nu_{a,IC}$ is of order the shock crossing time
$t_{\times}\sim T=50 T_{50}$ s. After the reverse shock crosses the
shell, electrons are no longer heated and the (on-axis) synchrotron flux 
at $\nu>\nu_a(T)$ drops, as does the the IC emission at $\nu >
\nu_{a,IC}$, and one starts to observe high latitude emission. 
The X-ray emission from the reverse shock decays steeply as $t^{-(p+4)/2}$,
whereas the forward shock emission decays $\propto t^{-(3p-2)/4}$.
Thus the slower decaying forward shock component eventually starts
to dominate.

\section{Dense ISM in Early Galaxies}

We consider a specific model in which the ISM density of early 
galaxies scales with redshift as $n \sim (1+z)^4 n_0$ cm$^{-3}$ 
(e.g. Ciardi \& Loeb, 2000). In this case (or in general when the ISM 
density is much larger than a typical $n_0\sim 1$ cm$^{-3}$ at $z=0$), 
a discussion analogous to that of the previous section can also lead to 
a bump in the reverse shock spectrum and in the optical/IR light curve.

The critical Lorentz factor classifying the evolution of the reverse
shock is given by (e.g. Kobayashi 2000) $\eta_{cr}^\prime 
\sim 90 ~n_0^{-1/8}\zeta_{6}^{-1/8}E_{53}^{1/8}T_{2.2}^{-3/8}$
where $\zeta_6=(1+z)/6$ and $T_{2.2}=T/150$ sec. A large fraction
of GRBs  are expected to be classified as thick shell cases, 
whose spectral quantities are given by 
(Kobayashi \& Zhang 2003a)
$\nu_c (T) \sim 2.0 \times 10^{11} 
n_0^{-1}\zeta_6^{-9/2}\epsilon_{B,-1}^{-3/2}
E_{53}^{-1/2}T_{2.2}^{-1/2}$ Hz, 
$\nu_m (T) \sim 7.2 \times 10^{11} n_0^{1/2}\zeta_6 \epsilon_{e,-2}^2
\epsilon_{B,-1}^{1/2} \eta_2^2$ Hz and
$F_{\nu,max} (T) \sim 27 n_0^{1/4}D^{-2}\zeta_6^{11/4}\epsilon_{B,-1}^{1/2}
E_{53}^{5/4}\eta_2^{-1}T_{2.2}^{-3/4}$ Jy, where 
$D=d_L(z)/(1.5\times10^{29} cm)$ is the normalized luminosity distance
and $D(z=5)=1$. Equating $F_{\nu_a}^{bb} \sim F_{\nu_a}$, we obtain 
$\nu_a (T) \sim 3.7 \times 10^{13} n_0^{1/7}\zeta_6^{3/14} 
\epsilon_{e,-2}^{2/7} \epsilon_{B,-1}^{1/14} 
E_{53}^{3/14}T_{2.2}^{-9/14}$ Hz. Using the scalings by Kobayashi (2000),
one can show that $\nu_a \propto t^{-3/7}$ during the shock crossing. 
The optical/IR light curve initially increases as
$\sim (\nu/\nu_a)^2 F_{\nu_a} \propto t^{9/7}$. When $\nu_a$ passes
through the K-band at $t_a \sim 5 ~ n_0^{1/3}\zeta_6^{1/2}
\epsilon_{e,-2}^{2/3}\epsilon_{B,-1}^{1/6}
E_{53}^{1/2} T_{2.2}^{-1/2} \nu_{obs,14.2}^{-7/3}$ sec, the flux reaches
a peak of $F_{\nu_K}(t_a) \sim 0.1 D^{-2}\zeta_6 \epsilon_{e,-2}
E_{53} T_{2.2}^{-1} \nu_{obs,14.2}^{-1}$ Jy, and then it rapidly
decreases by a (bump contrast) factor of $\mathcal{R}
\sim 30 ~n_0^{-1/4}\zeta_6^{-1/2} \epsilon_{e,-2}^{-1} 
\epsilon_{B,-1}^{-1/4} \eta_2^{-1} \nu_{obs,14.2}^{1/2}$ ($p=3$). 
After the emission drops to the level expected from the usual synchrotron 
model, the flux keeps a constant level,  and then it rapidly fades 
as $t^{-(p+4)/2}$ after a time comparable to the burst duration.

\section{Discussion and Conclusions}

We have analyzed the prompt afterglow emission from the reverse
shocks of GRBs occurring in dense environments, such as the stellar wind 
of a massive progenitor, or a dense ISM as might be expected in early galaxies. 
Usually in the fast cooling case,  
the $\nu F_{\nu}$ flux is normalized at the typical frequency $\nu_m$ 
by using the energy ejected into electrons. 
However, here we point out that if the synchrotron self 
absorption frequency $\nu_a$ is higher than the typical frequency 
$\nu_m$  (and $\nu_c$), this usual prescription should be
inapplicable for estimating the flux at and below $\nu_a$.
Such conditions can occur in the reverse shock of a
highly magnetized fireball in a dense environment. 
The radiation  flux suppressed by the self absorption effect is
redistributed among the electrons and photons in the optically 
thick regime, and most of it is emitted at $\sim \nu_a$ in a dynamical time. 
As a result, the self-absorption frequency is different and scales 
differently with the shock parameters, and the flux at the
self-absorption frequency shows a  bump which is a factor
$2(\nu_a/\nu_m)^{(p-2)/2} (\sim $ several) above the usual power law
flux estimate for typical parameters. The flux well above $\nu_{a}$ is
the same as before, but the flux below $\nu_{a}$ is larger by the same
factor $2 (\nu_a/\nu_m)^{(p-2)/2}$. 
This results in a new type of temporal behavior for the prompt optical  
flash of afterglows from fireballs in dense (e.g. wind  or early galaxy)
environments. These new features will be prominent when the inverse
Compton process is not important for electron cooling, i.e. for ejecta
with $\epsilon_e/\epsilon_B \ll 1$. This may be the case in fireball
ejecta which are highly magnetized, as suggested by some recent
studies. If, on the other hand, the burst occurs in a dense environment
and such features are absent, this may be an indication that
$\epsilon_e/\epsilon_B \gg 1$, and in this case a prompt reverse shock
X-ray flare is expected, which for a brief time dominates the forward
shock but decays faster than it.

Massive stars appear implicated in producing long gamma-ray bursts (as seen
from the detection of a supernova associated with GRB 030329, e.g., Stanek
et al 2003). Wind mass loss is expected from such stars previous to the
GRB explosion, but snapshot fits to forward shock late emission (e.g.,
Panaitescu \& Kumar, 2002) are compatible with such wind mass loss
in only a handful of cases. In general the parameters of stellar winds are 
poorly known, and the uncertainties are further increased at high redshifts, 
where massive stars are expected to be metal poor.  For this reason, 
signatures of a wind mass loss  or a dense environment would be extremely 
valuable, both for GRB astrophysics and for tracing the properties of
star formation at high redshifts. The prompt optical flashes expected
after tens of seconds from the reverse shock in a dense environment
would give  characteristic signatures in the spectral and temporal
behavior. These may help to test for the presence of winds and constrain
the wind mass loss, in moderate redshift environment, or alternatively,
at high redshifts they may provide evidence for a denser ISM than at low
redshifts. In such winds or dense ISM  the spectra and light curve time
behavior can also give constraints on the strength of the magnetic field
in the ejecta. 
 Large numbers of prompt X-ray detections with future missions such 
as {\it Swift}, complemented by ground-based follow-ups, should be able to 
test for such wind or dense ISM signatures and trace any changes with redshifts, 
if they exist, thus constraining the GRB environment as well as the radiation 
mechanisms.

We thank M.J. Rees, J. Granot, A. Beloborodov and  the referee for
valuable comments. This work is supported by NASA NAG5-13286, NSF 
AST 0098416, the Monell Foundation and the Pennsylvania State University 
Center for Gravitational Wave Physics, funded under cooperative agreement 
by NSF PHY 01-14375.


\clearpage

\end{document}